\documentclass[prd,showpcs,amsmath,amssymb,nofootinbib,longbibliography,twocolumn,showpacs]{revtex4-1}
\usepackage{multirow}
\usepackage{epsfig}
\usepackage{amsmath}
\usepackage{bm}
\usepackage{times}
\usepackage{graphicx}
\usepackage{color}
\usepackage{slashed}
\usepackage{graphicx}
\usepackage{amsmath}
\usepackage[latin1]{inputenc}
\usepackage{hyperref}
\usepackage{soul}

\def\bea{\begin{eqnarray}}
\def\eea{\end{eqnarray}}
\def\bean{\begin{equation*}}
\def\eean{\end{equation*}}

\begin{document}
 
\title{SU(5) Unification without Proton Decay}

\author{Bartosz~Fornal}
\affiliation{Department of Physics, University of California, San Diego, 9500 Gilman Drive, La Jolla, CA 92093, USA}
\author{Benjam\'{i}n~Grinstein}
\affiliation{Department of Physics, University of California, San Diego, 9500 Gilman Drive, La Jolla, CA 92093, USA}
\date{\today}

\begin{abstract}
  We construct a four-dimensional $\rm SU(5)$ grand
  unified theory in which the proton is stable. The Standard Model
  leptons reside in the $5$ and $10$ irreps
  of $\rm SU(5)$, whereas the quarks live in the $40$ and $50$ irreps. The
  $\rm SU(5)$ gauge symmetry is broken by the vacuum expectation
  values of the scalar  $24$ and $75$
  irreps. All non-Standard Model fields are heavy. Stability of the proton requires three relations
  between the parameters of the model to hold. However, abandoning the
  requirement of absolute proton stability, the model fulfills current
  experimental constraints without fine-tuning.\vspace{10mm}
\end{abstract}

\maketitle

\section{Introduction}\label{1} 
Grand unified theories (GUTs) present an attractive way to extend the Standard Model (SM) \cite{Glashow:1961tr,Weinberg:1967tq,Salam:1968rm,SU(3),Fritzsch:1973pi}. In addition to being esthetically appealing, they have several nice features -- they reduce the number of multiplets,  exhibit gauge coupling unification and explain why electric charges of quarks and leptons are connected.

The first attempt of partial unification was based on the group \,${\rm SU(4)} \times {\rm SU(2)}_L \!\times {\rm SU(2)}_R$ \cite{Pati:1974yy}, while the seminal papers describing full unification of couplings were those proposing $\rm SU(5)$  \cite{Georgi:1974sy} and $\rm SO(10)$  \cite{Fritzsch:1974nn} gauge groups. 
Unfortunately, GUTs with complete gauge coupling unification constructed so far in four dimensions are plagued with proton decay and the current experimental limit \cite{Miura:2016krn} excludes their simplest realization. Although there exist many models extending proton lifetime to an experimentally acceptable level (see \cite{Nath:2006ut} and references therein, including  orbifold GUTs), a theoretically interesting question remains:~is it at all  possible to construct a viable four-dimensional  GUT  based on a single gauge group with an absolutely stable proton?

In this letter we propose such a model. The main idea is simple but the
  realization is somewhat involved.  We present our model rather as a proof of concept,
anticipating a simpler realization in the future.  An alternative proposal achieves
  proton stability by imposing gauge conditions that eliminate all non-SM fields from the
  theory~\cite{Karananas:2017mxm}, resulting in a model that, however, appears to be
  indistinguishable from the SM.  The only other four-dimensional models with a single unifying
  gauge group designed to completely forbid proton decay we are aware of
  \cite{Segre:1980qc,Kuzmin:1981ek} are experimentally excluded due to the presence of new
  light particles carrying SM charges. 

The most dangerous proton decay channels in GUTs are those mediated by vector leptoquarks and arise
from gauge kinetic terms in the Lagrangian. In our model those channels are absent, since  the
quarks and leptons live  in different $\rm SU(5)$ representations. In particular, the leptons reside
in the $5$ and $10$ irreps of $\rm SU(5)$, the right-handed (RH) down quarks are formed
from a linear combination of two $50$ irreps, whereas the left-handed (LH) quark doublets
and the RH up quarks come from a linear combination of two $40$ irreps. The $\rm SU(5)$
gauge symmetry is spontaneously broken down to the SM by  vacuum expectation values (vevs) of
scalar field multiplets transforming as 24 and 75 irreps. In order to obtain correct SM masses,  the SM Higgs is chosen to be part of
a scalar  $45$ irrep multiplet, and there are no proton decay channels mediated by scalar leptoquarks from the Yukawa terms. 

The letter is structured as follows. In Sec.~\ref{2} we present the fermion and scalar content of the theory. Section~\ref{3} describes the relevant Lagrangian terms. In Sec.~\ref{4} we demonstrate that the SM fermions have SM Yukawa-type masses and all other fields in the theory are heavy. In Sec.~\ref{5} we show that proton decay is absent  at all orders in perturbation theory. We present conclusions and possible future directions in Sec.~\ref{6}.

\section{Particle content}\label{2} 
The model is based on the  gauge group $\rm SU(5)$. The fermion sector of the theory is composed of  the $5,$ $10,$ $40$ and $50$ irreps, where the $40$ and $50$ come in two vector-like copies, making the theory anomaly-free. The scalar sector consists of Higgs fields in the $24$, $45$ and $75$ irreps.

\subsection{Fermion sector}
The fermion multiplets in the theory come in the following LH spinor field representations, listed below along with their ${\rm SU(3)}_c \!\times {\rm SU(2)}_L \!\times{\rm U(1)}_Y$ decomposition \cite{Slansky:1981yr}:
\bea
&& \, 5^c \, = \ l \oplus D^c_{5} \nonumber\\ [3pt]
&&10 \ = \ e^c \oplus Q_{10}\oplus U^c_{10} \nonumber\\ [3pt]
&&40_{i} = Q_{40_i } \!\oplus U^c_{40_i } \!\oplus (1,2)_{-\frac{3}{2}}   \oplus (\bar{3}, 3)_{-\frac{2}{3}}  \oplus (8,1)_1 \oplus (\bar{6}, 2)_{\frac{1}{6}}   \nonumber\\[3pt]
&&\overline{{40}}_{{i}} = Q_{\overline{40}_{i}}^c\! \oplus {{U_{\overline{40}_i}}} \!\oplus (1,2)_{\frac{3}{2}}   \oplus ({3}, 3)_{\frac{2}{3}}  \oplus (8,1)_{-1} \oplus ({6}, 2)_{-\frac{1}{6}}   \nonumber\\[3pt]
&&50_{i}^c = D_{50_i}^c  \oplus(1,1)_{2}   \oplus (3, 2)_{\frac{7}{6}} \oplus (6, 3)_{\frac{1}{3}}  \oplus (\bar{6},1)_{-\frac{4}{3}} \nonumber\\
&&\hspace{10mm} \oplus \ (8, 2)_{-\frac{1}{2}}\nonumber\\[2pt]
&&\overline{{50^c_{i}}}= {D_{\overline{50}_i}}  \oplus(1,1)_{-2}   \oplus (\bar{3}, 2)_{-\frac{7}{6}} \oplus (\bar{6}, 3)_{-\frac{1}{3}}  \oplus (6,1)_{\frac{4}{3}}\nonumber\\
&&\hspace{10mm}  \oplus \ (8, 2)_{\frac{1}{2}} \ ,
\eea
where $i=1,2$. The lowercase fields $l, e$ are the LH lepton doublet and RH
electron, respectively. The fields  $Q, U$ and $D$ have the same quantum numbers as the SM's LH quark
doublet   $q$ and RH quark singlets $u$ and $d$, respectively.

When coupling to the $5^c$, ${\rm SU(5)}$ gauge bosons can act to transmute an $l$ to an anti-$D^c_5$,
  and when coupling to the 10 to transmute $Q_{10}$ to an anti-$U^c_{10}$. This is the standard
  route for proton decay in GUTs. If, however, the $5^c$ multiplet is split, in that the $D^c_5$
  mass is comparable to the GUT scale, while that of $l$ arises from electroweak symmetry breaking,
  and the light $d$ quark arises from a linear combination of the anti-$D^c_{50_i}$, then proton decay
  cannot proceed through this gauge boson exchange. This is an example of the realization of the
  mechanism we are proposing for proton stability.

\subsection{Higgs sector}
The scalar sector consists of the $24$, $45$ and $75$ irreps of $\rm SU(5)$. Their decomposition into SM multiplets:
\bea
24_H &=& (1,1)_0 \oplus (1,3)_0 \oplus (3,2)_{-\frac{5}{6}} \oplus (\bar{3},2)_{\frac{5}{6}} \oplus (8,1)_0 \nonumber\\ [2pt]
45_H &=& H \oplus (3,1)_{-\frac{1}{3}} \oplus (3,3)_{-\frac{1}{3}} \oplus (\bar{3},1)_{\frac{4}{3}} \oplus  (\bar{3},2)_{-\frac{7}{6}} \nonumber\\
&&\oplus \ (\bar{6}, 1)_{-\frac{1}{3}} \oplus (8,2)_{\frac{1}{2}}  \nonumber\\
75_H&=& (1,1)_0 \oplus (3,1)_{\frac{5}{3}} \oplus (\bar3,1)_{-\frac{5}{3}}\oplus (3,2)_{-\frac56} \oplus (\bar3,2)_{\frac56}  \nonumber\\
&& \oplus \ (\bar6,2)_{-\frac56}  \oplus  (6,2)_{\frac56} \oplus (8,1)_0 \oplus (8,3)_0 \ .
 \eea
 Only the Higgses in the $24$ and $75$ irreps develop  vevs at the
 GUT scale, which break the $\rm SU(5)$ gauge symmetry down to ${\rm SU(3)}_c \!\times \!{\rm SU(2)}_L
 \!\times \!{\rm U(1)}_Y$ \cite{Langacker:1980js,Hubsch:1984pg}. The SM Higgs field $H$ is part
 of the   $45$ irrep.

\section{Lagrangian}\label{3}
The fermion kinetic terms in the Lagrangian are:
\vspace{-3mm}

\bea\label{kin}
\mathcal{L}_{\rm kin}  &=& i \sum_R {\rm Tr}\left(\overline{R} \,\slashed{D}\, R\right) \, ,
\eea

\vspace{-1mm}
\noindent
where the sum is over the representations $R=$ $5^c$, $10$, $40_i$, $\overline{40}_{i}$, $50^c_i$ and $\overline{50^c_{i}}$. In the standard $\rm SU(5)$ GUT those terms give rise to dangerous dimension-six operators mediating proton decay. In our model such terms generating proton decay are absent, since physical states of SM quarks and leptons reside in different representations of $\rm SU(5)$, as  shown in Sec.~\ref{4}.

The Yukawa  interactions in our model are given by:
\bea\label{Lag5}
\mathcal{L}_Y & =&   Y_l \ 5^c 10 \,45^*_H  +    Y_{u}^{ij}\, 40_i\,40_j \,45_H + Y_d^{ij} 40_i\,50^c_j \,45^*_H\nonumber\\
&+&      M_{40}^{ij} \,\overline{40}_{i}\, {40_j} +  \lambda^{ij}_{1}  \,24_H \overline{40}_{i} \,{40_j} +  \lambda^{ij}_{2}  \, \overline{40}_{i} \,24_H{40_j} \nonumber\\[1pt]
& +&  \lambda_{3}^{i} \, 24_H 10 \, \overline{40}_{i} + \lambda^{ij}_{4}  \, \overline{40}_{i} \,75_H{40_j} + \lambda_{5}^{i} \,75_H 10 \, \overline{40}_{i}  \nonumber\\ [1pt]
& +& M_{50}^{ij} \,{50^c_i}\, \overline{50^c_{j}} + \lambda^{ij}_{6} \,{50^c_i} \,24_H \overline{50^c_{j}} + \lambda^{ij}_{7} \,{50^c_i} \,75_H \overline{50^c_{j}} \nonumber\\[1pt]
&+&   \lambda_{8}^{i}  \,75_H 5^c \, \overline{50^c_{i}} + {\rm h.c.} \ ,
\eea
with an implicit sum over $i,j=1,2$, the terms with $\lambda_{1,2}^{ij}$  corresponding to the two
independent contractions, and the Hermitian conjugate applied to non-Hermitian terms. In Eq.~(\ref{Lag5}) the coefficients of the only other allowed gauge-invariant renormalizable Yukawa terms ${Y'_u}^{i} 10\ 40_i \,45_H$ were set to zero.

  Since the SM leptons live only in the $5$ and $10$ irreps while the SM quarks live only in the $40$ and $50$ irreps, along with the absence of proton decay through
  vector gauge bosons, there is  no tree-level
  proton decay mediated by any of the Yukawa-type terms (contrary to other GUT models \cite{Dorsner:2012nq}). To see this, consider, for example,
  the first term in Eq.~\eqref{Lag5}: an exchange of the
$(3,1)_{-\frac13}$ of the $45$ necessarily couples the light
lepton doublet $l$ to the GUT-heavy $Q_{10}$. 

The Lagrangian of the scalar sector  consists of all possible renormalizable gauge-invariant terms involving the $24$, $45$ and $75$ representations:
\bea\label{scalarV}
\mathcal{L}_H&\! =\!& - \ \tfrac{1}{2}\mu_{24}^2{\rm Tr} (24_H^2)  \!+\!\tfrac{1}{4} a_1\!\!\left[{\rm Tr} (24_H^2) \right]^2 \!+\!\tfrac{1}{4} a_2{\rm Tr} (24_H^4) \nonumber\\[0pt]
&-& \tfrac{1}{2}\mu_{75}^2{\rm Tr} (75_H^2) \! +\!\tfrac{1}{4}\sum   b_k  {\rm Tr} (75_H^4)_k + M_{45}^2{\rm Tr} \big(|45_H|^2\big)\nonumber\\[0pt]
&+& \tfrac12 \sum g_k  {\rm Tr} (24_H^275_H^2)_k + \!\sum h_k  {\rm Tr} \big(24_H^2|45_H|^2\big)_k \nonumber\\
&+&  \ ... \ \ ,
\eea
where the index $k=1,2,3$ corresponds to the contractions in which the two  lowest  representations
in a given trace combine into a singlet, a two-component tensor and a four-component tensor,
respectively, and a prime is added if more than one contraction in each  case exists. For
simplicity, we exclude cubic terms in the scalar potential by assuming a $\mathcal{Z}_2$ symmetry of the Lagrangian.

\section{Particle Masses}\label{4}
In this section we show that  there exists a region of parameter space for which all SM fields have standard masses at the electroweak scale and below, whereas all new fields develop large masses.

\subsection{Fermion representations 5 and 50}
We first focus on the particles in the representation of the  down quark.~After $\rm SU(5)$ breaking, the corresponding Lagrangian mass terms are:
\bea
\mathcal{L}_{\rm mass} = \left(\begin{matrix} \, {D_{\overline{50}_1}} & {D_{\overline{50}_2}} \, \end{matrix}\right)
\mathcal{M}_D \left( \begin{matrix}
D^c_{5} \ \\
\,D^c_{50_1}\\
D^c_{50_2} 
\end{matrix} \right) ,
\eea
with the mass matrix elements
\bea
\begin{aligned}
{\mathcal{M}_D^{i,1}}&= \,\tfrac{\sqrt2}{3}\lambda_{8}^{i}v_{75} \ ,\\
{\mathcal{M}_D^{i,j+1}}&=\,{M_{50}^{ij}} + c^D_{24}\lambda_{6}^{ij}v_{24} + c^D_{75}\lambda_{7}^{ij}v_{75} \ ,
\end{aligned}
\eea
where $v_{24}$, $v_{75}$ are the vevs of the representations $24$, $75$, respectively, $c^D_{24} = 1/(3\sqrt{30})$ and $c^D_{75} = 1/(3\sqrt2)$.
In order to switch to the mass eigenstate basis, we perform a bi-unitary transformation
\bea
\mathcal{M}^{\rm diag}_D = ({R}_D)_{2\times2} \,\mathcal{M}_{D} \, ({L}_{D})^\dagger_{3\times3} 
\eea
and, correspondingly, the mass eigenstates are
\bea
\left( \begin{matrix}
{D^c_{5}}{'}\\
{D^c_{50_1}}\!\!\!\!'\ \\
{D^c_{50_2}}\!\!\!\!'\ \ 
\end{matrix} \right)_{\!\!L} \!\!={L}_D\!\left( \begin{matrix}
\!D^c_{5}\\
D^c_{50_1}\\
\!D^c_{50_2} 
\end{matrix} \right)_{\!\!L} \! , \  \ \ \left( \begin{matrix}
{D'_{\overline{50}_1}}  \\
{D'_{\overline{50}_2}} 
\end{matrix} \right)_{\!\!R} \!\!= {R}_D\!\left( \begin{matrix}
D_{\overline{50}_1}\\
D_{\overline{50}_2} 
\end{matrix} \right)_{\!\!R}. \ \  \ \ \ \ 
\eea 
The unitary matrices $L_D$ and $R_D$ are used to diagonalize the matrices $\left[ (\mathcal{M}_D)^\dagger \mathcal{M}_D\right]$ and $\left[\mathcal{M}_D (\mathcal{M}_D)^\dagger\right]$, respectively. 
From the structure of $\mathcal{M}_D$ we immediately infer that the matrix $\left[(\mathcal{M}_D)^\dagger \mathcal{M}_D\right]$ has one of the eigenvalues equal to zero. In order to completely forbid proton decay, the   corresponding eigenstate ${D^c_{5}}'$ cannot contain any admixture of $D^c_{5}$. This is achieved by requiring the following tuning of parameters\footnote{Condition (\ref{condition}) does not take into account terms involving the SM Higgs. With just this relation satisfied and no further fine-tuning of the electroweak terms, this would produce a tiny mixing between the heavy and light fields suppressed by $v/M_{\rm GUT}$, where $v$ is the SM Higgs vev and $M_{\rm GUT}$ is the unification scale. This would result in proton decay with lifetime $\tau_p \approx 10^{60} \ {\rm years}$. However, there exists a condition more general than (\ref{condition}) involving also the electroweak Yukawas, which ensures that there is no mixing between the SM quarks and the heavy fields. An alternative solution would be to stay with condition (\ref{condition}) and simply fine-tune $Y_{u}^{ij}$ and $Y_d^{ij}$, so that they produce exactly the SM quark mass terms, without any mixing between the light and heavy states.}:
\bea\label{condition}
{\rm det}\left({M_{50}^{ij}}+c^D_{24}\lambda_{6}^{ij}v_{24}  +c^D_{75}\lambda_{7}^{ij}v_{75}\right)  = 0 \ .
\eea
In this case ${D^c_{5}}'$  is a linear combination solely of $D^c_{50_1}$ and  $D^c_{50_2}$, and can be associated with the SM field $d^c$:
\bea\label{d}
d^c ={L}_D^{12} D^c_{50_1} +{L}_D^{13} D^c_{50_2} \ ,
\eea
where the matrix entries  ${L}_D^{1,j+1}$ are functions of $M_{50}^{ij}$,  $v_{24}$, $v_{75}$, $\lambda_6^{ij}$,  $\lambda_7^{ij}$ and $\lambda_8^i$.

{\renewcommand{\arraystretch}{1.4}\begin{table}[t!]
\begin{center}
    \begin{tabular}{| c | c | c |}
    \hline
       \ \ \ Field \ \ \ & \ \ $c_{24} \times \sqrt{30}$   \ \  &    \ \ $c_{75}  \times 3\sqrt2$ \ \   \\ [1pt]\hline\hline   
        \ \ \ \ \ $D^c_{50}$ \ \ \ \ \  \ & \ \ \ \ $1/3$ \ \ \ &  $\ 1$ \\[1pt] \hline 
 \ \ \ \ \ $(1,1)_{2}$ \ \ \ \ \  \ & \ \ \ \  $2$ \ \ \  & $\ 3$ \\[1pt] \hline      
  $ (3, 2)_{\frac{7}{6}}$ & $\ 7/6$ & $\ 1$  \\[2pt] \hline   
      $(\bar{6},1)_{-\frac{4}{3}}$ & $-\, 4/3$ & $\ 1$ \\ [2pt]\hline    
        $(6, 3)_{\frac{1}{3}} $ & $ \ 1/3$  &  \ \ \ $-\,1$ \ \ \  \\ [2pt]\hline  
        $(8, 2)_{-\frac{1}{2}}$ &  \raisebox{-0,3mm}{$-\,1/2$} & \raisebox{-0,3mm}{$\ 0$}  \\ [2pt]\hline   
        \end{tabular}
\end{center}
\vspace{-2mm}
\caption{\small{Contribution to the masses of the fermion components of the $50^c$ irrep generated by the Lagrangian terms in Eq.~(\ref{term}).}}
\label{tab01}
\end{table}}

The condition in Eq.~(\ref{condition}) ensures that our model has no proton decay that would involve either a component of the SM lepton doublet $l$ or the down quark $d$.
To our knowledge this novel model building feature has not been discussed in the literature.

If one chooses to abandon the requirement of absolute proton stability, the parameters of the model need not be tuned. Proton decay experimental constraints \cite{Miura:2016krn} require merely
\bea\label{re}
&&{L}_D^{11} \lesssim 0.1 \times \sqrt{({L}_D^{12})^2+({L}_D^{13})^2} \ . 
\eea

\vspace{-1mm}
\noindent
The factor of $\sim\!0.1$ can be easily understood: The presence of $D_5^c$ in ${D_5^c}'$ would trigger proton decay. The  standard $\rm SU(5)$ model predicts proton decay at a rate roughly 100 times larger than the current experimental bound. The contribution to this rate scales like the admixture of $D_5^c$ squared, thus the admixture itself has to be roughly less than 10\%.

Finally, one also has to show that all the fields within the $50^c$ irrep
other than $D_{50}^c$ are heavy. For this to be the case, it is sufficient to show that the Lagrangian terms:
\bea\label{term}
\Delta \mathcal{L}_{\rm mass} =  \lambda^{ij}_{6} \,{50^c_i} \,24_H \overline{50^c_{j}} +  \lambda^{ij}_{7} \,{50^c_i} \,75_H \overline{50^c_{j}}
\eea
generate different mass contributions:
\bea
\Delta \mathcal{M}^{ij} = c^R_{24}\lambda_{6}^{ij}v_{24} + c^R_{75}\lambda_{7}^{ij}v_{75}
\eea 
for those representations than for $D_{50}^c$, since then the equivalent of condition
(\ref{condition}) would not be fulfilled  for those representations and they would acquire GUT-scale
masses. The values of $c_{24}$ and $c_{75}$ are presented in Table~\ref{tab01}. When combined, these
    fulfill our requirements. Table~\ref{tab01} shows that the contribution of the term involving
        the $75$ irrep in Eq.~(\ref{term})  gives the same mass
        for   $D_{50}^c$  as for $ (3, 2)_{\frac{7}{6}}$  and
        $(\bar{6},1)_{-\frac{4}{3}}$. The contribution of the term involving
        the  $24$ irrep  in Eq.~(\ref{term}) breaks this degeneracy.

{\renewcommand{\arraystretch}{1.4}\begin{table}[t!]
\begin{center}
    \begin{tabular}{| c | c | c | c |}
    \hline
       \ \ \ \ Field \ \ \ \ & \ \ $c_{24_1}\times \sqrt{30}$  \ \  &  \ \ $c_{24_2}\times \sqrt{30}$   \  \ &  \ \  $c_{75}\times 3 \sqrt2$ \  \ 
       \\ \hline\hline   
\ \ \ \ $U^c_{40}$ \ \ \ \ & $\ 13/9 $ & $\ 1/3$ & $\ 5/9$ \\[2pt] \hline   
\ \ \ \ $Q_{40}$ \ \ \ \ & $-\,7/9$ &$-\, 4/3 $  & $\ 1/9$ \\[2pt] \hline          
\ \ \ $(1,2)_{-\frac{3}{2}}$ \ \ \ & $\ 2 $ & $-\,3$ & $\ 1 $  \\[2pt] \hline     
  $ (\bar{3}, 3)_{-\frac{2}{3}} $ & $ \ 1/3$ & $-\, 3$ & $-\, 1/3$ \\[2pt] \hline 
      $(\bar{6}, 2)_{\frac{1}{6}} $ & $\ 1/3 $ &$\ 2$ & $-\, 1/3$ \\ [2pt]\hline     
      $(8,1)_1$ & $-\,4/3 $ & $\ 2$ & $\ 1/3$ \\ \hline
        \end{tabular}
\end{center}
\vspace{-1.5mm}
\caption{\small{Mass contribution generated by the terms involving the scalar $24$ and $75$ for the fermion components of the $40$ irrep.}}
\label{tab02}
\end{table}}

\subsection{Fermion representations 10 and 40}
The analysis for the ${\rm SU(3)}_c \!\times \!{\rm SU(2)}_L \!\times\!{\rm U(1)}_Y$ representations with the quantum numbers of the quark doublet $Q$ and anti-up quark $U^c$ is a little different, since they both reside in the  $40$  of $\rm SU(5)$. Following the reasoning from the previous case, we arrive at the two conditions:
\bea\label{conditionQ}
&&{\rm det}\!\left[{M_{40}^{ij}}\!  +\!\big(c^{U,Q}_{24_1}\lambda_{1}^{ij}\!+\! c^{U,Q}_{24_2}\lambda_{2}^{ij}\big)v_{24}\!+\! c^{U,Q}_{75}\lambda_{4}^{ij}v_{75}\right]  = 0 \ , \ \ \ \ \ \  \ \ 
\eea
with the values of the coefficients provided in Table~\ref{tab02}.
If these relations are fulfilled, the SM fields $u^c$ and $q$ are not part of  the $10$ irrep,  preventing the proton from decaying through channels involving $q$, $u$ and $e$. We verified that there exists a class of values for the parameters $M_{40}^{ij}$, $\lambda_{1,2,4}^{ij}$ fulfilling the requirement (\ref{conditionQ}),  thus forbidding proton decay. The SM $u^c$ and $q$ are given by:
\bea\label{uq}
\begin{aligned}
u^c&= L_U^{12} \,U^c_{40_1} + L_U^{13} \,U_{40_2}^c \ ,\\
q\,&= L_Q^{12} \,Q_{40_1} \!+ L_Q^{13} \, Q_{40_2} \ ,
\end{aligned}
\eea
where $L_{U, Q}^{1, j+1}$ are  functions of $M_{40}^{ij}$, $v_{24}$, $v_{75}$, $\lambda_{1,2,4}^{ij}$ and  $\lambda^i_{3,5}$.

\noindent
The values of $c_{24_1}^R$, $c_{24_2}^R$ and $c_{75}^R$ for the other
${\rm SU(3)}_c \!\times {\rm SU(2)}_L \!\times {\rm U(1)}_Y$ components of the $40$ are
given in Table~\ref{tab02}. All those representations have different
sets of $c^R$'s as compared to $U^c$ and $Q$  and, consequently,
  Eq.~(\ref{conditionQ}) is not satisfied in those cases. Therefore,
 those
representations develop GUT-scale masses.

\subsection{Scalar representations 24, 45 and 75} 
In our model the gauge group $\rm SU(5)$ is broken down to the SM by the GUT-scale
  vevs of the $24$ and $75$ irreps, while the $45$ does not develop a vev. Stability of
  the scalar potential is equivalent to the condition that all squared masses of the components of the $24$ and $75$ irreps are
 positive, except for one combination of $(3,2)_{-\frac{5}{6}}$ and one of
  $(\bar3,2)_{\frac{5}{6}}$ \cite{Langacker:1980js,Hubsch:1984pg,Cummins:1985vg}, the would-be Goldstone bosons of the
  broken $\rm SU(5)$.  We checked that there exists a large region of parameter space for
  which all components of the $24$ and $75$ develop large positive squared masses,
  apart from the $(3,2)_{-\frac{5}{6}}$ and $(\bar3,2)_{\frac{5}{6}}$ for which the
  mass-squared matrix is given by
\bea\label{matrix32}
\mathcal{M}^2_{{(3,2)}} = -\tfrac{1}{18} (g_2+11 \,g_3+ 15\,g_3')   \left( \begin{matrix}
\frac{v_{75}^2}{5} &  \frac{v_{24} v_{75}}{2\sqrt{10}}\\
 \frac{v_{24} v_{75}}{2\sqrt{10}}&   \frac{v_{24}^2}{8}
\end{matrix} \right)\!. \ \ \ \ \ \ \ 
\eea
We have used relations between parameters satisfied at the stationary point of the potential. The constant of proportionality is a combination of coupling constants, defined in Eq.~(\ref{scalarV}),  and can take either sign. The matrix (\ref{matrix32})
 has a vanishing determinant so that one of the linear combinations of the fields is massless while  the other is heavy.

The representation $45$ does not take part in $\rm SU(5)$ breaking and its
${\rm SU(3)}_c \!\times {\rm SU(2)}_L \!\times{\rm U(1)}_Y$ components generically have
masses at the GUT scale. Since one of those fields is the SM Higgs, a cancellation between
some of the parameters of the potential is required. To show that such an arrangement is
possible, it is sufficient to consider only the  explicit mass term
for the $45$ along with the terms mixing it with the $24$ in Eq.~(\ref{scalarV}). A small SM
Higgs mass contribution is obtained for: 
\bea\label{M45} 
M_{45}^2 + \left(h_1 -
  \tfrac{67}{240} h_2 + \tfrac{31}{120}h_2'- \tfrac{13}{60} h_3 - \tfrac{5}{16}h_3'
\right) v_{24}^2 \simeq 0 \,. \ \ \ \ \ \ \ 
\eea 
We verified that there exists a wide
range of parameters for which the GUT-scale masses of all other components of the $45$ are
positive.  The fine-tuning in Eq.~(\ref{M45}) is equivalent to the standard $\rm SU(5)$
doublet-triplet splitting problem and perhaps may be solved by introducing additional
$\rm SU(5)$ representations along the lines of \cite{Grinstein:1982um,Masiero:1982fe}.

\subsection{Quark and lepton masses}
The SM electron Yukawa emerges from the term:
\bea
Y_l \,5^c 10 \,45^*_H \supset  y_l \,l\, H^* e^c \ .
\eea
The terms contributing to the SM down quark mass are:
\bea\label{third}
Y_d^{ij}40_i\,50^c_j \,45^*_H \supset y_d \,q \, H^* d^c \ ,
\eea
and for the SM up quark we have:
\bea\label{fourth}
Y_{u}^{ij} 40_i\,40_j \,45_H  \supset  y_u \,q\, H\, u^c  \ .
\eea
There is no need to correct the typical $\rm SU(5)$ relation between the electron and down quark Yukawas, since they are not directly related in our model. 

\section{Proton stability at loop level}\label{5}
So far, we have shown that the model proposed in this letter is completely free from any tree-level proton decay. As it turns out, it is also possible to forbid proton decay at any order in perturbation theory. 

First we note that the model has  no proton decay at any loop order mediated by vector gauge bosons or scalars from the $45$ irrep. This can be argued on symmetry grounds. All the Lagrangian terms in Eqs.~(\ref{kin}) and (\ref{Lag5}), apart from  $\lambda_{3}^{i} \, 24_H 10 \, \overline{40}_{i}$, $\lambda_{5}^{i} \,75_H 10 \, \overline{40}_{i}$ and  $ \lambda_{8}^{i}  \,75_H 5^c \, \overline{50^c_{i}} $, are invariant under:
\begin{align}\label{transf}
5^c \rightarrow - 5^c \ ,& \ \  10 \rightarrow -10 \ . \ 
\end{align}
Under this transformation, the SM leptons are odd while the SM quarks are even.  For proton decay one must have an odd number of leptons in the final state and none in the initial state, and there must be no heavy particles in either the initial or final states. This is odd under the transformation  (\ref{transf}), and hence forbidden.

The only remaining loop-level proton decay channels are those mediated by the scalars from the $24$ and $75$ irreps. To forbid these, we assume that the spontaneous breaking of ${\rm SU}(5)$ is nonlinearly realized \cite{Coleman:1969sm} and we can replace the $24$ and $75$ irreps by nondynamical condensates \cite{Karananas:2017mxm}. The $24$ and $75$ scalar sector of the theory is then described by a nonlinear sigma model \cite{GellMann:1960np,Callan:1969sn}. This concludes the proof that  in our model the proton is stable.

\section{Conclusions}\label{6}
We have constructed a grand unified model in four dimensions based on the gauge group $\rm SU(5)$ which does not exhibit any proton decay. This was accomplished by assigning the quarks and leptons to different irreps of $\rm SU(5)$. In order to forbid proton decay at tree level, three relations between the model parameters have to hold. In addition, for proton stability at any loop order, the ${\rm SU}(5)$ breaking has to be nonlinearly realized.
Abandoning the requirement of absolute proton stability removes the necessity of any tuning or the nonlinear symmetry breaking, and the model is consistent with experiments for a large range of natural parameter values.

The model has  additional desirable features.  Upon adding one \cite{Stone:2011dn} or several \cite{Murayama:1991ah,Cox:2016epl} extra scalar representations it allows
for gauge coupling unification if some of the scalar fields from the $45$ irrep are at the  TeV scale. It also contains no problematic relation between the electron and down quark Yukawa plaguing the standard $\rm SU(5)$ models.
However, the usual doublet-triplet splitting problem still persists and requires further model building, perhaps along the lines of a non-supersymmetric version of \cite{Grinstein:1982um}.

Let us stress again that our goal was just to show through an explicit
construction that, contrary to common belief, four-dimensional grand
unified theories with a stable proton do exist.~We hope that this may inspire new directions in model building  and
revive the interest in grand unification, which  perhaps deserves
more attention in spite of negative results from proton decay
experiments.

\subsection*{Acknowledgments}
We are grateful to Ilja Dor\v{s}ner and the anonymous Physical Review Letters referees for constructive comments regarding our manuscript. This research was supported in part by the DOE Grant No.~${\rm DE}$-${\rm SC0009919}$. \vspace{10mm}

\bibliography{SU5_revised}

\end{document}